\documentclass[10pt, a4paper]{article}

\usepackage{lrec-coling2024} % this is the new style
\usepackage{booktabs} % For horizontal rules
\usepackage{xcolor}
\usepackage{makecell}

\title{An Automated End-to-End Open-Source Software for High-Quality Text-to-Speech Dataset Generation}

\name{Ahmet Gunduz, Kamer Ali Yuksel, Kareem Darwish, Golara Javadi \\ 
      {\bf \large Fabio Minazzi, Nicola Sobieski, Sébastien Bratières}}
\address{aiXplain, Inc. \\
         \{ahmet, kamer, kareem, golara\}@aixplain.com \\ \\
         Translated, Inc. \\
         \{fabio, nicola, sebastien\}@translated.com}

\abstract{
Data availability is crucial for advancing artificial intelligence applications, including voice-based technologies. As content creation, particularly in social media, experiences increasing demand, translation and text-to-speech (TTS) technologies have become essential tools. Notably, the performance of these TTS technologies is highly dependent on the quality of the training data, emphasizing the mutual dependence of data availability and technological progress. This paper introduces an end-to-end tool to generate high-quality datasets for text-to-speech (TTS) models to address this critical need for high-quality data. The contributions of this work are manifold and include: the integration of language-specific phoneme distribution into sample selection, automation of the recording process, automated and human-in-the-loop quality assurance of recordings, and processing of recordings to meet specified formats. The proposed application aims to streamline the dataset creation process for TTS models through these features, thereby facilitating advancements in voice-based technologies. \newline \Keywords{Text-to-Speech Dataset Generation, Automated Recording and Quality Assurance}
}

\begin{document}

\maketitleabstract

\section{Introduction}

The advent of artificial intelligence (AI) has brought about transformative changes across a multitude of industries, from healthcare and finance to entertainment and communication. One of the most notable areas impacted by AI is voice-based technologies, which have seen significant advancements in recent years. Text-to-Speech (TTS) systems, in particular, have become increasingly sophisticated, finding applications in various sectors, including assistive technologies, content creation, and customer service. The recent achievement with data-centric approaches proved to be as crucial as model architecture \cite{mazumder2022dataperf}. Especially for complex models, the data quality stands out as much as the quantity of it. Therefore, developing high-quality TTS models is contingent upon the availability of comprehensive and well-curated datasets. This process often involves labor-intensive tasks such as data selection, recording, and annotation.

Generating datasets suitable for TTS models involves several steps, each with its challenges. Sample selection, for instance, must consider the complete  coverage of phonemes to make it possible for the resultant TTS model to reproduce all the sounds of the target language. The recording process, too, requires meticulous planning and execution to guarantee the audio quality is up to par. Furthermore, the assurance of recording quality often necessitates using Automatic Speech Recognition (ASR) models to validate the data. Finally, preprocessing steps are essential to convert the raw recordings into a format amenable to training.

Given these complexities, we introduce an integrated tool to streamline the dataset generation process for training TTS models, which reduces manual effort and enhances the quality and reliability of the datasets produced. Our proposed open-source tool is unique because it enables the rapid preparation of text for recording, batch processing of recorded audio, and quality assurance.  To our knowledge, no other integrated open-source tool with similar functionality is available. The contributions of this work are multifaceted, namely:% by introducing an end-to-end application to generate high-quality TTS datasets (Figure \ref{fig:workflow}).

\begin{itemize}
    \item A novel approach for sample selection that diversifies language-specific phoneme distribution, thereby enhancing the linguistic richness of the dataset.
    \item An automated recording process that minimizes human intervention, increasing efficiency and  enabling the speakers to focus on the voice performance.
    \item Quality assurance mechanisms powered by ASR models are integrated into the system to validate the recording accuracy and quality.
    \item Preprocessing functionalities that prepare the recordings for subsequent model training.
\end{itemize}
The code  used in this work is publicly available.\footnote{\url{https://github.com/aixplain/tts-qa}}

\begin{figure*}[ht]
    \centering
    \includegraphics[width=\textwidth]{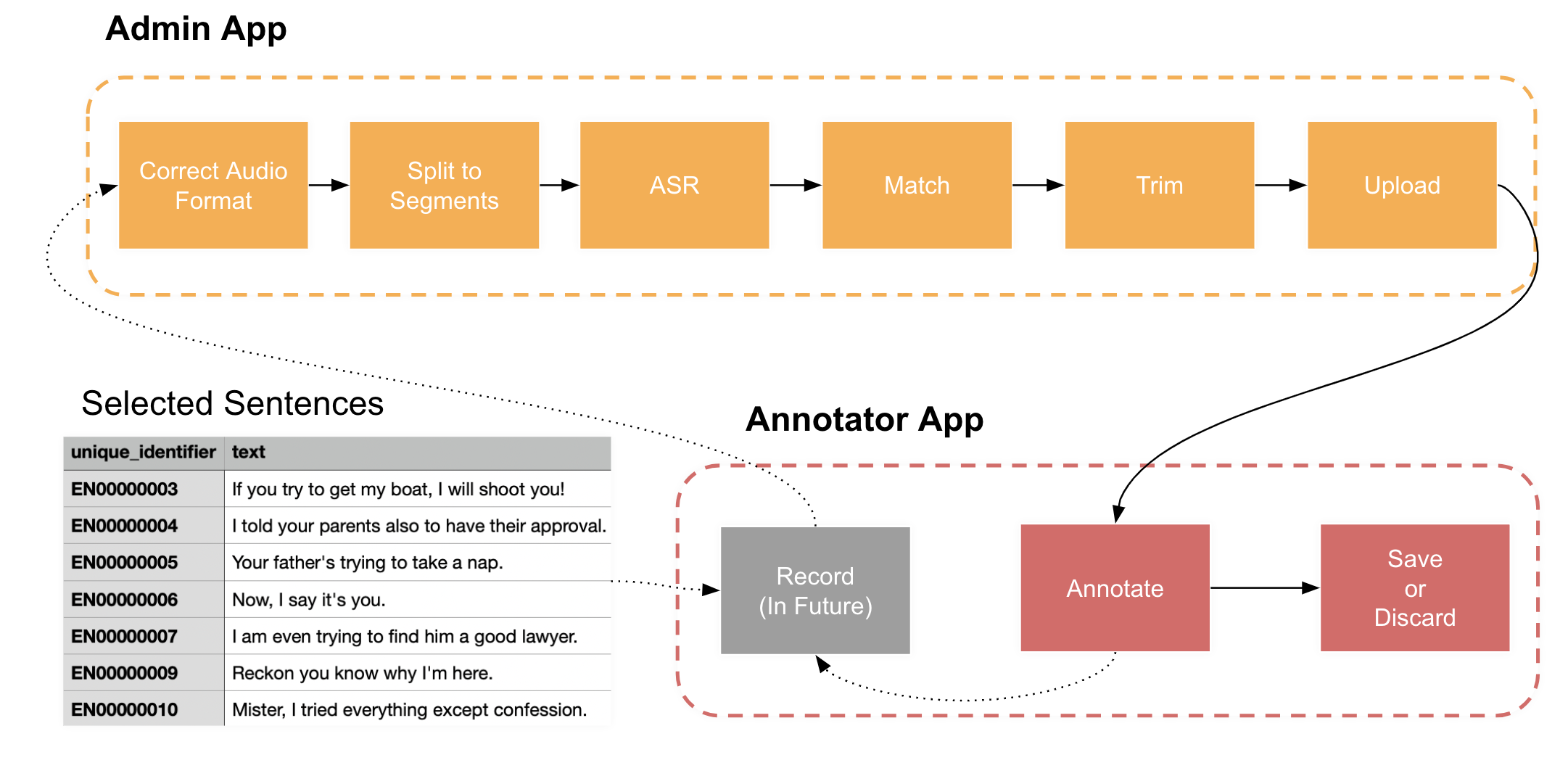} 
    \caption{Workflow of the System}
    \label{fig:workflow}
\end{figure*}

\section{Related Work}
Text-to-Speech (TTS) has seen significant advancements in recent years, primarily due to the application of deep learning techniques \cite{jeong2021diff,zhang2023complete}. Various architectures, such as Tacotron \cite{wang2017tacotron} and WaveNet \cite{oord2016wavenet}, were proposed to improve the naturalness and intelligibility of synthesized speech. 

Following the development of renowned generative modeling frameworks, like Generative Adversarial Networks and Normalizing Flows \cite{goodfellow2014generative,rezende2015variational}, their application in TTS engines became prominent. These frameworks facilitated parallel generation, ensuring the quality of synthesized speech remained consistent. After the WaveNet paper's release in 2016, there was a surge in efforts to develop a parallel non-autoregressive vocoder for high-quality speech synthesis. Architectures such as Parallel WaveNet and WaveGlow \cite{prenger2019waveglow,oord2018parallel}, rooted in Normalizing Flows, not only accelerated the inference process but also upheld superior synthesis quality, especially evident on GPU devices \cite{popov2021grad}.

Recently, a new category of generative models called Diffusion Probabilistic Models (DPMs), has demonstrated its proficiency in modeling intricate data distributions, encompassing areas like images, shapes, graphs, and handwriting. The fundamental concept of DPMs revolves around a two-step process: Initially, a forward diffusion process is constructed by progressively deconstructing the original data until a basic distribution is achieved. Subsequently, a reverse diffusion, parameterized by a neural network, is designed to trace the paths of the forward diffusion in reverse time. two vocoders representing the DPM family showed impressive results in raw waveform reconstruction: WaveGrad \cite{chen2020wavegrad} and DiffWave \cite{kong2020diffwave} were shown to reproduce the fine-grained structure of human speech.

However, the quality of these models is highly dependent on the datasets used for training. Previous works have explored different aspects of dataset creation, including data augmentation techniques, phoneme-based selection, and quality assurance through manual annotation or semi-automated methods. Similarly, in machine translation (MT), approaches to dataset generation to optimize the annotation process have been adopted \cite{yuksel-etal-2022-efficient}. This paper introduces a system with comprehensive features designed to expedite the generation of datasets for TTS applications.

\section{Data and Preprocessing}

Data were sourced from publicly available repositories, specifically the OPUS corpus\footnote{\url{https://opus.nlpl.eu}}. Six languages were targeted for this study: German, English, Mandarin, Italian, French, and Spanish. The datasets across different languages contained raw sentences (segments from the dataset not subjected to processing or cleaning). The study aimed to generate 30 hours of audio recordings for each language. Several text and audio constraints were established to ensure the data quality.

The initial step involved the extraction of sentences from the scripts collected for each target language. Subsequently, a filtering process was applied to these sentences to eliminate those containing numerical values or abbreviations, so that the dataset can be used independently from the normalization rules that will be adopted in the training and inference phases. Additionally, sentences were filtered based on predetermined length constraints to account for the typical attention span of the current TTS engines. The sentences that met these criteria were fed into our text analysis system for further processing. Tables \ref{tab:text_data_criteria} and \ref{tab:audio_data_criteria} outline the specific criteria set for the text and audio samples.

\begin{table*}[th]
    \centering
    \small % Reduce font size
    \begin{tabular}{c|m{12cm}}
        \toprule
        \textbf{Criteria} & \textbf{Description} \\
        \midrule
        Sentence Types & The dataset should encompass a variety of sentence structures, including declarative (80\%), interrogative (10-15\%), and exclamatory (5-10\%) sentences, as indicated by ending with period, question mark, and exclamation mark, respectively. \\
        % \midrule
        % Punctuation & Sentences should incorporate a diverse set of punctuation marks, including but not limited to commas, semicolons, periods, colons, and ellipses. \\
        \midrule
        Sentence Length & Each sentence should be no fewer than five and no more than 13 words. \\
        % \midrule
        % Sentence Distribution & Approximately 80\% of the sentences should be declarative, 10-15\% should be interrogative, and 5-10\% should be exclamatory. \\
        \midrule
        Normalization & The dataset should exclude acronyms, abbreviations, digits, or symbols necessitating text normalization. \\
        % \midrule
        % Abbreviations & Sentences should not contain abbreviations. \\
        % \midrule
        % Pattern Diversity & Sentences should not exhibit repetitive patterns in terms of words or phrases. \\
        \bottomrule
    \end{tabular}
    \caption{Text Data Criteria}
    \label{tab:text_data_criteria}
\end{table*}

\begin{table*}[th]
    \centering
    \small % Reduce font size
    \begin{tabular}{c|m{12cm}}
        \toprule
        \textbf{Criteria} & \textbf{Description} \\
        \midrule
        File Format & WAV, Mono channel \\
        \midrule
        Sampling Rate & 88 kHz \\
        \midrule
        Sample Format & 16-bit, PCM \\
        \midrule
        Peak Volume Levels & from -3 dB to -6 dB \\
        \midrule
        Signal-to-Noise Ratio & Not less than 35 dB \\
        \midrule
        Silence Duration & Leading and trailing silences should not exceed 100 ms; internal silences should not exceed 0.5 seconds. \\
        \midrule
        Audio Artifacts & The recordings should be free from lip-smacking, echo, and breath sounds. \\
        \midrule
        Recording Length & Each recording should be no longer than 15 seconds and no shorter than 2 seconds. \\
        \midrule
        Speech Rate & Recordings should be made at a natural speed. \\
        \midrule
        Accent & The accent in the recordings should align with the target language. \\
        \midrule
        Punctuation Accuracy & The audio should accurately reflect the punctuation in the text. \\
        \bottomrule
    \end{tabular}
    \caption{Audio Data Criteria}
    \label{tab:audio_data_criteria}
\end{table*}

\begin{figure*}[t]
    \centering
    \includegraphics[width=2\columnwidth]{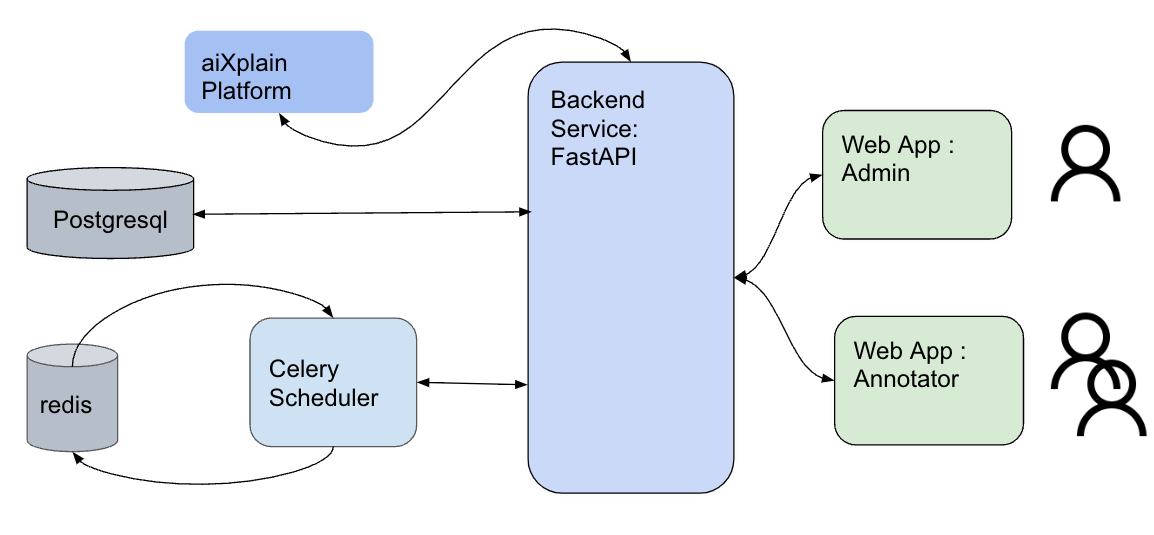} % Replace 'your_figure_filename' with the actual file name
    \caption{System Architecture of the Tool}
    \label{fig:system_architecture}
\end{figure*}

\section{Methodology}
The methodology section provides an in-depth description of the proposed end-to-end tool for TTS dataset generation. The tool is designed to be modular, allowing for customization at each stage of the dataset creation process. Figure \ref{fig:workflow} indicates the overall workflow of the system. The first module focuses on sample selection, employing language-specific phoneme distribution algorithms to ensure a balanced and comprehensive dataset. The second automates the recording process, providing a user-friendly interface for capturing high-quality audio samples. The third leverages Automatic Speech Recognition (ASR) models for quality assurance, flagging recordings that do not meet predefined quality criteria. The final module preprocesses the recordings into formats compatible with existing TTS training pipelines.

The system employs the {aiXplain SDK}\footnote{\url{https://docs.aixplain.com/main.html}} and platform\footnote{\url{https://platform.aixplain.com}} for model-specific tasks such as Automatic Speech Recognition (ASR) and Voice Activity Detection (VAD). Streamlit is utilized for the user interface due to its straightforward design and ease of implementation. Each service is containerized using Docker, and the entire project can be deployed using a Docker Compose file for reliability.

\subsection{Sample Selection}
Within the text analysis system, the sentences underwent a phonemization process to capture the distinct phonetic elements inherent to each language. For this purpose, the Espeak Phonemizer\footnote{\url{https://github.com/rhasspy/espeak-phonemizer}} was employed to generate monophones, diphones, and triphones for each sentence. After this, frequencies of the generated phonemes were computed for each sentence and aggregated into a dictionary object, forming an initial corpus-level distribution of phonetic elements. This served as a foundational phonetic reference for each language.

\begin{figure}[ht]
    \centering
    \includegraphics[width=\columnwidth]{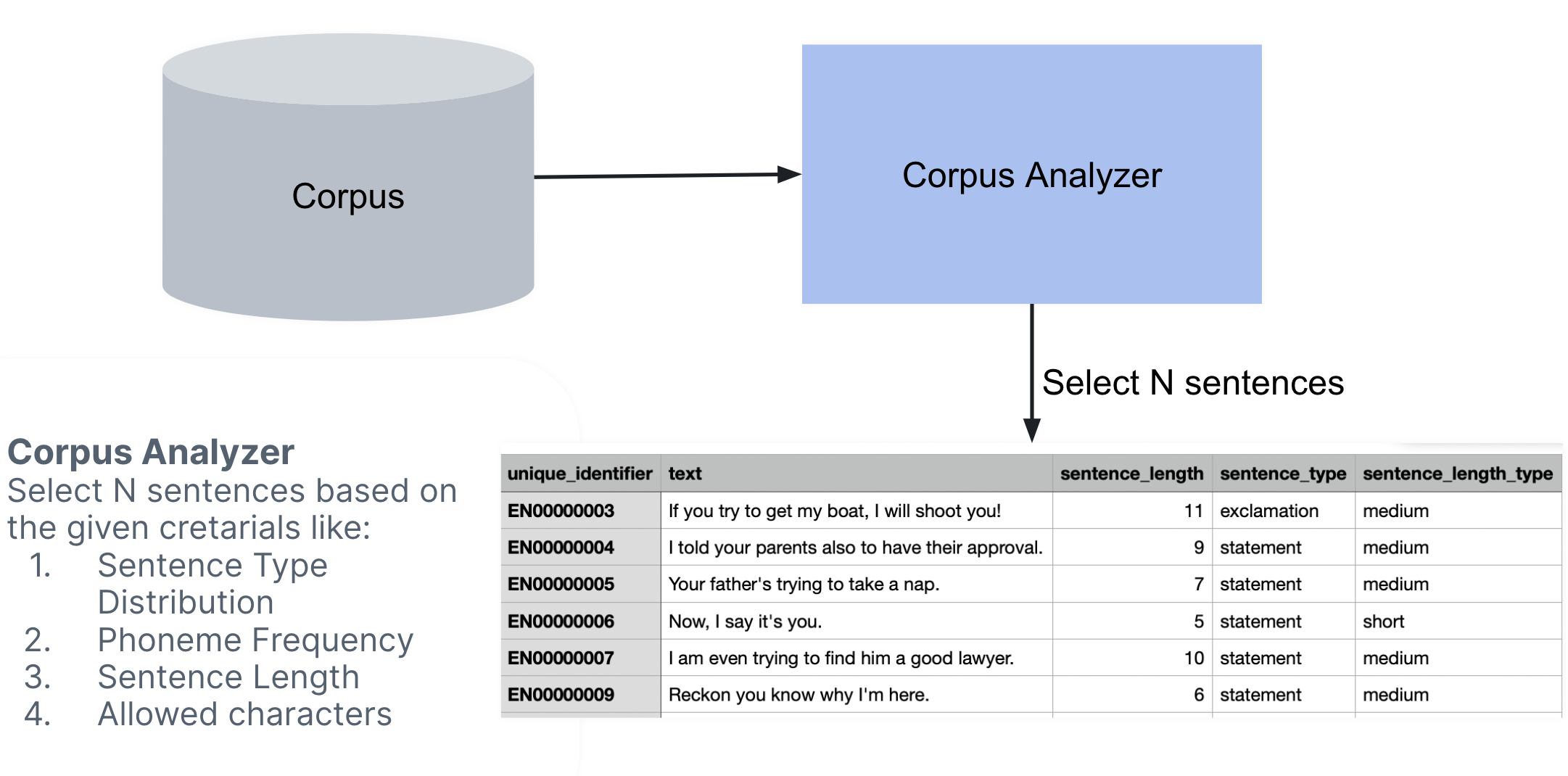} 
    \caption{Sentence Selection for TTS Recordings}
    \label{fig:sentence_selection}
\end{figure}

To construct a representative subset of the corpus, an iterative selection process was employed, guided by multiple criteria: the type of sentence, the length of the sentence, and the phonetic distribution relative to the corpus-level distribution. The algorithm for sentence prioritization employs stochastic selection, but assigns higher probability values to sentences that contribute to aligning the subset's phonetic distribution more closely with that of the overall corpus. This alignment is quantified by calculating the divergence between the phonetic distribution of the current subset and the corpus-level distribution. In this iterative process, sentence type and length constraints are applied as filters to guide the selection of subsequent samples, ensuring that the resulting subset remains within predefined percentage boundaries for these criteria. Figure \ref{fig:sentence_selection} shows the workflow of the sentence selection process. Each sentence was weighted according to the frequency of its constituent phonemes. Sentence selection for the final dataset was performed through iterative sampling to align the selected sentences' phonetic elements with the corpus-level distribution. This selection process was initialized randomly and refined iteratively.

To quantify the dataset, a target of a minimum of 600,000 words per language was set. Utilizing a reference rate of 2.75 words per second, it was determined that this word count would be sufficient to produce more than 30 hours of audio recordings.

\subsection{Preparing Recordings}\label{sec:prepare_recording}
Voice actors were permitted to utilize their preferred audio recording and editing software for self-directed recording sessions, provided they adhered to the criteria specified in Table \ref{tab:audio_data_criteria}.  We also gave the voice actors the choice of saving the recording of each sentence in a separate file or batch recording one sentence after another into one file.  If a voice actor decides to record in batch, they need to adhere to the following guidelines:
\begin{itemize}
    \item A single file should have a maximum of 500 sentences.
    \item File naming should follow the "start\_ID-end\_ID.wav" convention (e.g., DE00000037-DE00000720.wav).
    \item A minimum of 2 seconds should be maintained between each sentence.  This is essential to perform accurate Voice Activity Detection (VAD) for identifying sentence audio segments.
    \item In case of errors, the voice actor can re-read a sentence (as many times as (s)he likes), with the condition that the last iteration is correct.
\end{itemize}

The batch recording is initially split into segments using VAD. Each segment is then transcribed using an ASR, and the Levenshtein edit distance is computed between the ASR output and all the sentences between the start and end IDs in the recording name. Initially, Whisper\footnote{\url{https://github.com/openai/whisper}} was employed as the default ASR system, owing to its language-agnostic capabilities. However, due to the auto-correction features inherent to Whisper, which led to discrepancies in the dataset, it became necessary to transition to alternative ASR systems available on the aiXplain platform.  Though we used Azure, one of the advantages of using the aiXplain platform is that we can swap in and out different ASR systems with no changes to the code. The sentence with the lowest edit distance is picked given if: a) the ratio of edit distance to the minimum length of ASR output and sentence is less than 0.2, and the difference in length between the ASR output and the sentence does not exceed 20\% of the length of the shorter of the two. Given these conditions, we achieved an average match rate of 99\%. Subsequently, matched segments are trimmed using VAD to remove leading and trailing silences exceeding 100 ms.  We also ensured there was at least 25 ms of silence to avoid any speech truncation.

\subsection{Automated Processing of Recordings}
This module aims to help with data creation for text-to-speech (TTS), which involves two primary components: the user interface and the back-end, which facilitate audio data annotation and storage.

\subsubsection{User Interface}
\begin{figure}[ht]
    \centering
    \includegraphics[width=\columnwidth]{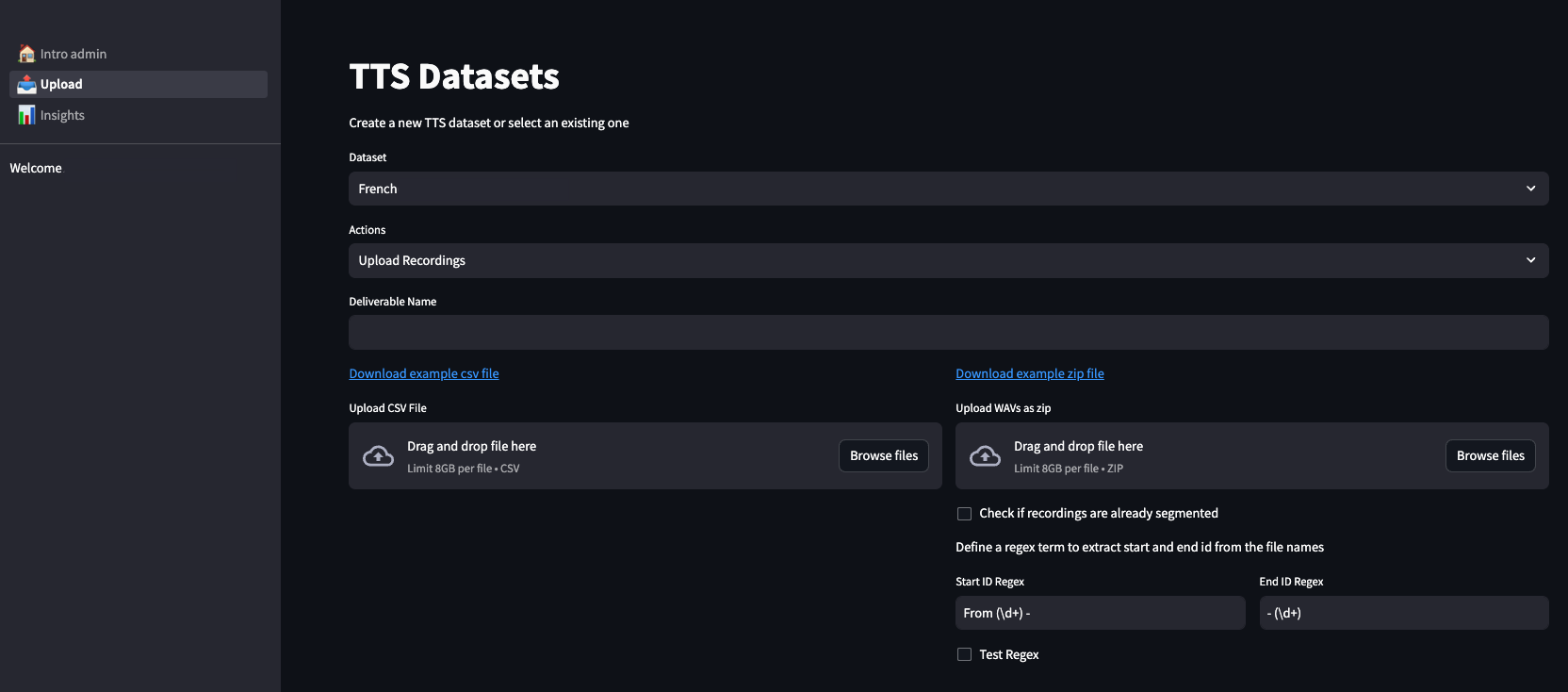} 
    \caption{Screenshot of Admin Dashboard for Recording Upload}
    \label{fig:upload}
\end{figure}

The user interface is bifurcated into an Admin Dashboard and an Annotator Dashboard, as depicted in Figures \ref{fig:upload} and \ref{fig:annotation}, respectively. For privacy considerations, some user-specific information has been redacted from these figures.

\paragraph{Admin Dashboard:}
This interface is responsible for various administrative tasks such as uploading recordings from voice actors, dataset creation, visualization of annotation progress, user account management, and task assignment.

\paragraph{Annotator Dashboard:} 
This interface is designed to annotate recordings.  As can be seen in the screenshot, annotation involves multiple tasks: 
\begin{itemize}
    \item Verifying that the audio matches the corresponding sentence and or editing the text as necessary to ensure they match entirely. The annotators can post-edit the original text or the ASR output.
    \item Marking recordings as having specific problems such as repetitions and incorrect prosody. 
\end{itemize}
More details about annotations are provided in the Quality Assurance section (Section \ref{section:quality_assurance}).

\subsubsection{Backend}
The system architecture, illustrated in Figure \ref{fig:system_architecture}, comprises a PostgreSQL database, a Celery scheduler\footnote{\url{https://docs.celeryq.dev/en/stable/}}, and a Redis backend orchestration service and in-memory database\footnote{\url{https://redis.io/}}. Using Celery and Redis helps manage various asynchronous tasks, such as audio segmentation and running ASR. % and quality for optimizes time efficiency of the upload and processing tasks, the Celery framework is employed for scheduling, while Redis is used for asynchronous operations. Recordings may be submitted in one of two formats: segmented or batch.
The processed and annotated recordings are stored in the PostgreSQL database and S3 buckets. Both raw and processed audio files are also saved to local directories to ease subsequent processing.

\begin{figure}[ht]
    \centering
    \includegraphics[width=\columnwidth]{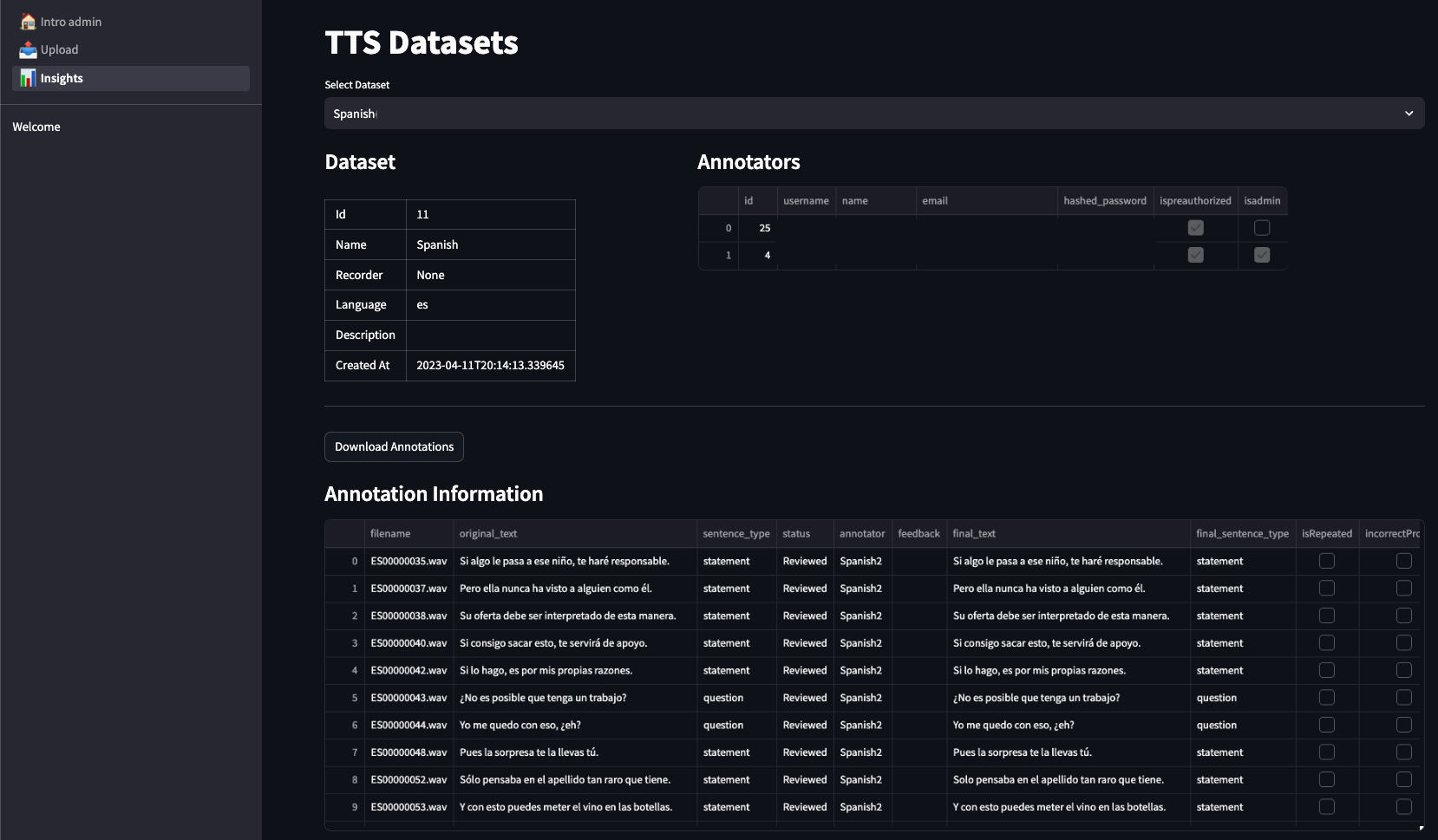} 
    \caption{Screenshot of Admin Dashboard for Insights of the annotations done.}
    \label{fig:insight}
\end{figure}

\subsection{Quality Assurance}
\label{section:quality_assurance}
\begin{figure}[ht]
    \centering
    \includegraphics[width=\columnwidth]{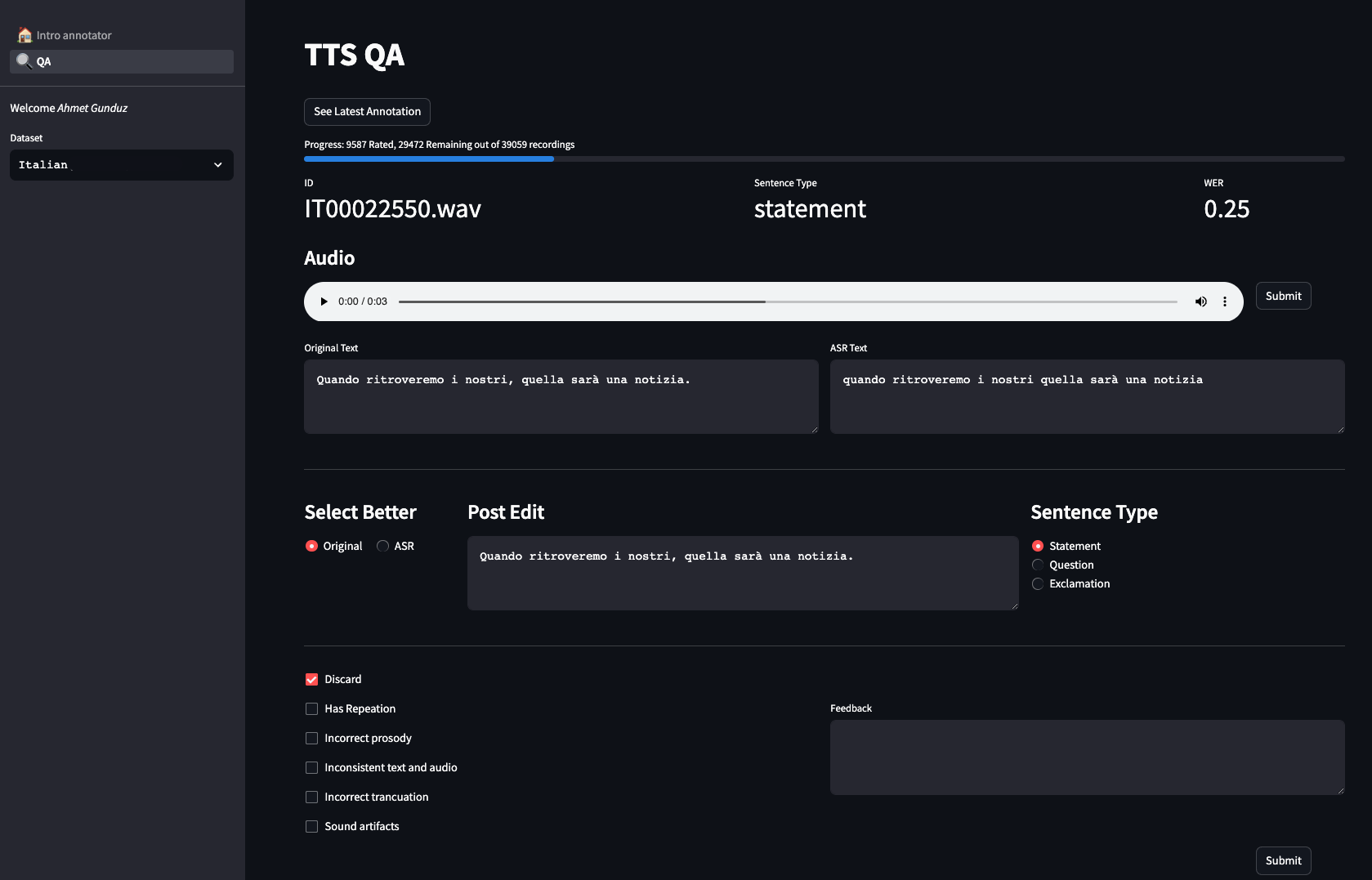} 
    \caption{The Screenshot of Annotator App}
    \label{fig:annotation}
\end{figure}

After the initial preprocessing and uploading of audio recordings, the first Quality Assurance is performed by a team of annotators native in each target language. The annotation phase ensures that each recording meets the criteria for high-quality samples. Figure \ref{fig:annotation} displays a screenshot of the annotation interface. Administrators, utilizing the Admin Dashboard, allocate specific datasets to individual annotators. A user-level authentication system monitors annotator activities and adds a layer of security. Access is restricted to accounts created using pre-authorized email addresses.

Upon logging in, annotators are presented with the datasets assigned to them. Individual recordings are sequentially retrieved for annotation when a dataset is selected, prioritizing those with the highest Word Error Rate (WER), as computed by the ASR system against the reference text. Annotators can listen to the audio, make text edits to ensure exact alignment with the audio (including pauses for punctuation), modify the sentence structure, and discard recordings for various reasons. These reasons may include word or phrase repetition, improper prosody, inconsistency between audio and text, incorrect pronunciation, or auditory artifacts such as noise or lip-smacking. An additional feedback field is available for annotators to elucidate the rationale behind discarding a sample. % Upon completing the sample review, annotators can submit their annotations, at which point a new sample with the highest WER is presented, continuing the process until no more samples remain.

The annotation system is designed to accommodate multiple annotators working concurrently, even on identical datasets. Each sample is locked to a specific annotator session to prevent overlap, ensuring that annotators do not conflict. Due to the resource-intensive nature of the annotation process, only a single annotation is permitted for each sample. Moreover, administrators can oversee the annotation process via the Admin Dashboard, which features statistical data on the Insight page. This functionality allows administrators to monitor each task's progress and maintain the highest possible annotation quality. As illustrated in Figure \ref{fig:insight}, administrators can review all datasets and download annotated data for subsequent TTS training.

\begin{table*}
    \centering
    \small % Reduce font size
    \begin{tabularx}{\textwidth}{l|l|X|X|X|X|X|X|X}
        \toprule
        Lang & File & Dur. Before Match. & Dur. After Match. & Dur. After Trim. & Total Files & Assigned & Not Assigned & \% Assigned \\
        \midrule
        DE & File1 & 2439.02 & 1549.47 & 1330.77 & 495 & 480 & 15 & 97.0\% \\
        DE & File2 & 2354.78 & 1493.00 & 1274.15 & 494 & 486 & 8 & 98.4\% \\
        FR & File1 & 2271.79 & 1465.28 & 1241.03 & 498 & 491 & 7 & 98.6\% \\
        FR & File2 & 2326.11 & 1475.08 & 1253.37 & 498 & 488 & 10 & 98.0\% \\
        ES & File1 & 2505.61 & 1499.82 & 1286.45 & 498 & 491 & 7 & 98.6\% \\
        ES & File2 & 2216.55 & 1464.50 & 1241.68 & 500 & 488 & 12 & 97.6\% \\
        IT & File1 & 2249.54 & 1473.51 & 1247.73 & 496 & 489 & 7 & 98.6\% \\
        EN & File1 & 1906.00 & 1285.45 & 1020.80 & 530 & 503 & 27 & 94.9\% \\
        EN & File2 & 2692.67 & 1241.19 & 1011.56 & 500 & 499 & 1 & 99.8\% \\
        \bottomrule
    \end{tabularx}
    \caption{Performance Metrics of Sentence Matching Algorithms Across Multiple Languages and Files. The table summarizes the duration before and after matching, the duration after trimming, and the number of sentences assigned and not assigned. The percentage of sentences successfully assigned is also included, with performance generally exceeding 98\%, except in one English recording where it is 94.9\%.}
    \label{tab:file_summary_by_language}
\end{table*}

\section{Experimental Setup and Results}
In this section, we discuss the experiments conducted to evaluate the efficiency and quality of the trimming process. To evaluate the efficacy of the proposed tool, a series of experiments were conducted using multiple languages: German (DE), French (FR), Spanish (ES). Italian (IT), and English (EN). % Two tables summarize the key metrics: 
Table \ref{tab:dataset_summary} provides an overview of the datasets before and after annotation, while Table \ref{tab:file_summary_by_language} offers a detailed breakdown of file durations, sentence assignments, and assignment percentages by language in our matching algorithm. One of the notable observations from Table \ref{tab:file_summary_by_language} is the reduction in total duration after trimming silences from the audio files. For instance, in the German dataset, the duration decreased from 1549.47 seconds to 1330.77 seconds after trimming. This reduction makes the dataset more compact and enhances its quality by eliminating non-informative silences.

\subsection{Matching Efficiency}

The matching efficiency, as observed from the "Dur(ation) Before Match." and "Dur(ation) After Match." columns in Table \ref{tab:file_summary_by_language}, is approximately 98\% across multiple languages. This high efficiency indicates that most audio data aligns well with the corresponding text, ensuring the dataset is highly reliable for further processing as machine learning.

Moreover, the high percentage of assigned sentences, as indicated in the last column of Table \ref{tab:file_summary_by_language}, further corroborates the efficiency of the matching algorithm. For example, in the English dataset, 94.9\% to 99.8\% of the total segments were successfully assigned. As this has been observed for all languages, we conclude that the method's accuracy will persist if the voice actors follow the requirements mentioned in Section \ref{sec:prepare_recording}. 

\subsection{Quality of Annotated Data}

The annotation process plays a crucial role in ensuring the dataset's quality. As seen in Table \ref{tab:dataset_summary}, the percentage of post-edited samples is relatively low, such as 1.25\% for the Spanish dataset. Additionally, the rate of discarded samples is negligible, standing at 0.00\% for the same dataset. These low percentages suggest that most of the data is of high quality and requires minimal intervention during the annotation phase.

\subsection{Second Review Pass}

To ensure the result of the overall process is not affected by the individual annotators, we included a second review step by another set of annotators native to the respective target languages. We reviewed at least 70\% of the datasets, not all of them, due to time and budget restrictions.
The instructions given to the second set of annotators were the same as those provided for the first review step.
The quality of the edits was carefully screened to verify whether the annotators involved in the first Quality Control stage correctly interpreted the instructions, especially regarding matching the speech and the punctuation. Proper training of TTS engines requires a close matching between punctuation and speech, considering that punctuation is an essential cue for neural networks to correctly interpret subordinate and coordinate sentences, exclamations, and questions. 
The results in Table 5 show that the toolset developed for this study effectively generates high-quality data.

\begin{table*}
    \centering
    \small % Reduce font size
    \begin{tabular}{c|c|c|c|c|c|c|c}
        \toprule
        Language &  \# of Samples & Bad Prosody &  \makecell{Inconsistent \\ Text-Audio} & Truncation & \makecell{Sound \\ Artifacts} & \% Edited & \% Discarded \\
        \midrule
        German  &  30000 & 0 & 1 & 0 & 21 & 1.90\% & 0.15\% \\
        Spanish &  45489 & 0 & 0 & 0 & 0 & 1.25\% & 0.00\% \\
        Italian &  30001 & 0 & 0 & 0 & 0 & 11.38\% & 0.00\% \\
        English &  33373 & 2 & 0 & 3 & 0 & 1.44\% & 0.02\%  \\
        French  &  30005 & 0 & 0 & 0 & 0 & 3.23\% & 0.00\% \\
        \bottomrule
    \end{tabular}
    \caption{Performance metrics for the second stage of Quality Control Across Multiple Languages. The table provides the total number of samples, the type of errors found, the \% of segments edited, and the \% of segments discarded in this second QC.}
    \label{tab:dataset_summary}
\end{table*}

\subsection{Review of the Recordings}

The annotation process primarily focuses on reviewing the recordings, which are often the most critical in determining the overall quality of the dataset. The high percentages of assigned sentences and the low percentages of post-edited or discarded samples suggest that the end recordings generally meet the quality requirements, yielding a high-quality dataset suitable for various applications. In summary, the experiments demonstrate that the annotation process is highly efficient at creating high-quality datasets. The trimming of silences and the high matching efficiency contribute to the efficiency of the recording process, while the meticulous annotation process ensures its quality. Italian stands as an exception in this study as the speaker could not correctly perform the script, so in both QC phases, many adjustments had to be made. 

\subsection{Staff for TTS dataset collection}

One of the results of this study is that to create a high-quality dataset for TTS the process outlined here has to be operated by a team of professionals, each native in one of the target languages. The speakers should be able to control their prosody according to the punctuation and intonations indicated in the brief and the script. Also, it should be apparent what the expected inflection of the speech should be e.g., a speaker with a regional cadence is not desirable for building a dataset aimed at training models that have to speak the official language of a country. The annotators should also be native, as they must be able to catch inflections and flaws in prosody, which can confuse the TTS engines at training time with low correlations between text and speech.
This is especially true for small training datasets (<100h), where the statistical error cancellation has a lower impact than large datasets.

\section{Discussion and Limitations}

This paper presents a comprehensive tool for generating high-quality Text-to-Speech (TTS) datasets applicable across various languages. However, there are inherent limitations concerning the tool's reliance on high-quality Automatic Speech Recognition (ASR) models for quality assurance. Such models may not be universally available for all languages or dialects. Initially, the Whisper ASR model was employed with its default configurations, serving as a language-agnostic solution. Nonetheless, the auto-correction functionality within Whisper posed challenges in accurately segmenting the audio recordings. Specifically, during the batch recording process, voice actors are instructed to repeat a sentence if an error occurs while recording. If insufficient pauses are made between these repetitions, the Voice Activity Detection (VAD) system interprets the repetitions as a single sentence recording. Subsequently, Whisper's auto-correction feature alters the sentence transcripts, leading to inaccuracies in the dataset. To mitigate this issue, we resorted to utilizing other ASR services on aiXplain platform to cross-verify and correct recordings.

Additionally, for segmented recordings, there were instances where voice actors or actresses incorrectly labeled the recordings with erroneous sentence identifiers. To rectify this issue, we applied our matching algorithm -initially designed for batch recording uploads— to all segmented recordings. Specifically, the algorithm cross-referenced the original sentence associated with the filename against the ASR-generated transcript of the recording. Re-matching was conducted using the Levenshtein edit distance, employing the same methodology as in the batch recording matching process. As a consequence of these challenges, we intend to facilitate recording samples directly through our proprietary tool in future iterations. This approach aims to minimize the likelihood of labeling errors and streamline the overall data collection process.

\section{Future Work}

Future work could focus on several avenues to enhance the tool's capabilities. The first is developing an efficient recording tool to minimize the speakers' distraction from their performance. In light of the challenges encountered with erroneous labeling of segmented recordings, we intend to facilitate recording samples directly through our proprietary tool in future iterations. This approach aims to minimize the likelihood of labeling errors and streamline the overall data collection process. Another potential direction is the integration of more advanced ASR models to improve the quality assurance process, especially for under-represented languages. Another avenue could be incorporating machine learning algorithms to automate the annotation process further, thereby reducing the need for human intervention. Additionally, the tool could be extended to support more complex data types and formats, making it more versatile and applicable to a broader range of TTS applications.

To further enhance the quality and accuracy of the generated TTS dataset, an innovative approach can be employed by leveraging an unreferenced speech dataset like movies or video recordings using a reference-less metric, such as NoRefER \cite{yuksel23_icassp, yuksel23_interspeech}, which enables the assessment of transcription accuracy without the need for a reference transcription. This capability is particularly beneficial in creating high-quality TTS datasets from speech data that has not been previously transcribed or for which no reliable reference exists. Selecting these high-fidelity transcriptions for the TTS dataset ensures a foundation of exceptional quality. Incorporating suggestions from analyzing the NoRefER attentions can streamline curating a high-quality dataset by improving the annotation efficiency and effectiveness \cite{javadi2024wordlevel}. 
 
\section{Conclusion}

This paper has presented an end-to-end tool engineered to automate and streamline the creation of high-quality datasets for Text-to-Speech (TTS) models. To our knowledge, no other tool with similar capabilities currently exists in the literature or the market. The tool incorporates several innovative features, including a sample selection algorithm for language-specific phoneme distribution, an automated recording process, and quality assurance mechanisms powered by Automatic Speech Recognition (ASR) models. Experimental results across multiple languages demonstrate the tool's efficacy in producing datasets that are both comprehensive and of high quality. The annotation process, facilitated by a user-friendly interface, further ensures the reliability of the generated datasets.

The proposed tool represents a significant step forward in TTS dataset generation, offering a scalable and efficient solution for creating high-quality datasets. Its modular design allows for easy customization and adaptation, making it a valuable resource for researchers and practitioners alike in the rapidly evolving landscape of voice-based technologies. One of the key advantages is the tool's ability to automate and streamline the dataset creation process, thereby reducing the time and effort required. 
However, it is essential to acknowledge certain limitations. The overall dataset quality depends on the professional level of the team operating it. The tool's quality assurance mechanisms rely on the availability and accuracy of ASR models, which may not be universally applicable across all languages or dialects. Additionally, while the tool significantly reduces the manual effort required in dataset creation, human intervention is necessary.

\nocite{*}
\section{References}\label{sec:reference}

\bibliographystyle{lrec-coling2024-natbib}
\bibliography{lrec-coling2024}

\end{document}